\documentclass[10pt]{iopart}
\usepackage[T1]{fontenc}
\usepackage{graphicx}
\usepackage{url}
\usepackage{hyperref}
\usepackage{times}
\usepackage{enumitem}
\newcommand{\mM}{\rm M}
\newcommand{\me}{\rm e}
\newcommand{\mC}{\rm C}
\newcommand{\mO}{\rm O}
\newcommand{\mH}{\rm H}

\begin{document}

\title[C$_n$H electron attachment effects on C$_n$H$^-$/C$_n$H ratio]
  {Modeling the role of electron attachment rates on column density ratios for C$_n$H$^-$/C$_n$H ($n=4,6,8$) in dense molecular clouds}

\author{F.A. Gianturco}
\ead{francesco.gianturco@uibk.ac.at}
\address{Institut f\"ur Ionen Physik und Angewandte Universit\"at Innsbruck, Technikerstrasse 25, A-6020, Innsbruck, Austria}

\author{T. Grassi}
\address{Center for Stars and Planet Formation, Niels Bohr Institute, University of Copenhagen, \O stervoldgade 5-7, DK-1350 Copenhagen \O, Denmark}

\author{R. Wester}
\address{Institut f\"ur Ionen Physik und Angewandte Physik, Universit\"at Innsbruck, Technikerstrasse 25, A-6020, Innsbruck, Austria}

\begin{abstract}

 The fairly recent detection of a variety of anions in the Interstellar Molecular Clouds have underlined
  the importance of realistically modeling the processes governing their abundance.
   To pursue this task, our earlier calculations for the radiative electron attachment (REA) rates for C$_4$H$^-$,
C$_6$H$^-$, and C$_8$H$^-$ are employed in the present work, within a broad network of other
 concurrent reactions, to generate the corresponding column
density ratios of anion/neutral (A/N) relative abundances . The latter are then  compared
 with those  obtained in recent years from observational measurements.
The calculations involved the time-dependent solutions of a large network of
chemical processes over an extended time interval and included a series of runs
in which the values of REA rates  were repeatedly scaled over
several orders of magnitude. Macroscopic parameters for the Clouds' modeling were 
also varied to cover a broad range of physical environments.
 It was found that, within the range and quality of the processes
included in the present network,and selected from state-of-the-art astrophysical databases, 
 the REA values required to match the $observed$ A/N
ratios needed  to be reduced by orders of magnitude for C$_4$H$^-$ case, while the same
rates for C$_6$H$^-$ and C$_8$H$^-$ only needed  to be scaled  by much smaller factors.
The results suggest that the generally proposed  formation of interstellar anions by REA 
mechanism is  overestimated by current models for the C$_4$H$^-$ case, for which is 
likely to be an inefficient path to formation . This path is thus providing  a
 rather marginal contribution to the observed abundances of C$_4$H$^-$, the latter being 
more likely to originate from other chemical processes in
the network, as we discuss in some detail in the present work.
Possible physical reasons for the much smaller differences against observations found instead  
 for the  values of the (A/N) ratios in two other, longer members of the series 
are put forward and analyzed within the  evolutionary modeling discussed in the present work.

\end{abstract}

\submitto{\jpb}

\maketitle

\ioptwocol

\section{Introduction}
\label{intro} 

 Molecular negative ions in the Interstellar Medium (ISM) were first
discovered as an unidentified series of lines within  a radio astronomical survey
 of the evolved carbon star IRC+10216 by \cite{Kawaguchi1995} and were later assigned by \cite{McCarthy2006}, who confirmed the 
presence of C$_6$H$^-$ in TMC-1 on the basis of laboratory rotational spectroscopy.
The observation was followed by the detection of both C$_6$H$^-$ and
C$_4$H$^-$ in the protostellar core L1527 by \cite{Sakai2007} and by \cite{Agundez2008}.
The following year \cite{Gupta2009} carried out an extensive dedicated
survey of C$_6$H$^-$ in twenty four different molecular sources: they
observed that anion only in two star-forming clouds and confirmed the
anion-to-neutral (A/N) ratio of the order of a few percent, in keeping with the
earlier measurements.

 Those detections had been limited to the Taurus Molecular Cloud
complex until the more recent discoveries of the presence of the C$_6$H$^-$
anion in the Lupus, Cepheus and Auriga star-forming regions \cite{Sakai2010,Cordiner2011}
which therefore established the widespread presence of
molecular anions in different corners of the ISM.
The presence of similar molecular species, the cyanopolyynes, has also been
observed in the same star-forming cloud IRC-10216 via the detection of
C$_3$N$^-$ by \cite{Thaddeus2008}.

 The possible chemical importance of molecular anions in the ISM had 
been put forward over the years even well before their recent detections.
The work in \cite{Dalgarno1973}, in fact,  first discussed this possibility suggesting the
chief mechanism of formation to be the Radiative Electron Attachment (REA)
path: M + e$^-\rightarrow$ M$^-$ + $h\nu$.
Attempts at detection using radioastronomy were discussed by \cite{Sarre1980}, while
\cite{Herbst1981} further argued that large interstellar molecules, including the
polyynic chains, could efficiently undergo the anionic stabilization process
via the REA process.
Evolutionary models were later put forward, after the anions' observations of  above, 
by \cite{Millar2007} and by \cite{Harada2008},
although marked discrepancies were found between the estimated A/N ratios for
C$_4$H$^-$ and C$_6$H$^-$ and those suggested by the above models \cite{Agundez2008}.
It has also been demonstrated later on by \cite{Cordiner2012} that anion
measurements have the potential to offer additional information on the
molecular properties of interstellar clouds because of the marked reactivity of
these ionic species, the latter being very sensitive to the relative
abundances of electrons and of C, H, N atoms in the environment.

 It therefore becomes important to be able to model the evolutions of
the anionic abundances in different molecular clouds in order to
find out which are the most important molecular mechanisms that are likely to preside
over their chemical formation .

 In the present work we shall therefore revisit the A/N ratios
recently observed experimentally in the dense interstellar medium \cite{Cordiner2013} by 
using the existing estimated rates  of REA processes
which have been obtained by  calculations. We shall employ such rates 
for generating the A/N ratios and the latter quantities will  be compared with those  from
 astronomical observations, a comparison that  will help us to evaluate what is the
 expected importance of REA anions' formation vis \'{a} vis other chemical routes 
currently present in the  existing databases and routinely used within chemical 
evolutionary models. In our study we shall refer to a  database that we here label
 as KIDA\footnote{\url{http://kida.obs.u-bordeaux1.fr/}}, \cite{Wakelam2015}.

 The following two sections briefly review the existing computational 
models for the REA anions' formation and further discuss our use of them in the
evolutionary calculations we have employed  to produce the A/N ratios over large time intervals.
Such calculations involve an extended network of chemical processes that have to be 
 included  to  realistically describe the  environment of a dark cloud core.
Our final results and our present conclusions are given in Section~\ref{sect:conclusions}.

\section{The molecular steps of REA}

 In order to gain a better physical insight into the process of anion
stabilization by photon emission, frequently considered the most likely
mechanism in the ISM and originally put forward to explain their formation by various authors
\cite{Dalgarno1973,Sarre1980,Herbst1981}, an extensive modelling
for polyatomic species was discussed, and applied to both polyynes and
cyanopolyynes by \cite{Herbst2008}. Here we shall call this treatment the Herbst-Onamura (HO) model for 
the REA  mechanism.
More recently, a different  model based on more realistic scattering calculations was put
forward by \cite{Carelli2013} who found in general similar results for the
electron attachment rates to polyynes as those suggested by \cite{Herbst2008},
although uniformly smaller than the latter over the whole range of temperature
examined in both studies.Here we shall call this model the Carelli-Satta-Grossi-Gianturco 
(CSGG) modelling of REA processes of anion formation.

 Both models essentially consider the presence of two different, albeit
cooperating, steps in  the stabilization of the final molecular
anions:

 (i) the direct photon emission during the electron attachment
process, whereby the incoming electron is taken to have rather little excess
collision energy, finds the neutral radical in the ISM to be largely in its
ground vibrational normal modes ($\bar{\nu}$ $\sim$ 0) and the emitted photon 
releases the necessary excess energy to form the corresponding anion in its ground
vibrational mode ($\bar{\nu}'$ $\sim$ 0) without fragmentation of the target

\begin{equation}
 \mM(\bar{\nu} \sim 0) + \me^- \rightarrow \mM^-(\bar{\nu}' \sim 0) + h\nu\,.
\end{equation}

 Both models take this process to occur in competition with the
autodetachment event, without explicit inclusion of the
latter decay process within their computational treatments.

 (ii) The indirect electron stabilization path, whereby the incoming
projectile is temporarily trapped by the neutral target into a metastable state of the
corresponding anion:

\begin{equation}
 \mM(\bar{\nu} \sim 0) + \me^- \rightarrow \mM^{-*}(\bar{\nu}' \ne 0) + h\nu\,.
\end{equation}

 Here  the above two models treat differently the  non-adiabatic
(electron-molecule) couplings within the metastable anionic complex:

 (i) The HO model uses a phase-space theory (PST) approach whereby strong coupling exists 
among the  initial state of the anion and any of the excited states of the latter with the
same spin. The REA efficiency increases with the size of the radical and
the number of its normal modes which are energetically available.

 (ii) The treatment  of the CSGG model explicitely includes the
occurrence of either low-energy shape resonances or of intermediate dipole
driven states near the scattering threshold (above and below it) for the systems with
supercritical dipoles ($\mu>2.0$~Debye). Such metastable states increase the efficiency of
radiative stabilization vs autodetachment and the model found that the REA paths  become more 
competitive with autodetachment as the size of the molecular system increases \cite{Carelli2013} .

 One should note that in both models, once the bound anionic molecules are formed, 
 the competing reaction of  photodetachment , which can be
induced by the environmental photons, is also present:

\begin{equation}
 \mM^{-}(\bar{\nu}' \sim 0) + h\nu \rightarrow \mM(\bar{\nu} \ne 0) + \me^{-}\,.
\end{equation}

 On the other hand, in  modelling the anions' formation processes in the core of the dense
molecular clouds one  expects marginal penetration by photons.
It has been noted, in fact, that essentially no molecular anions were observed
in the PDR regions \cite{Agundez2008}, thereby confirming  the
important role of process (3) in those external regions but its reduced importance in
the  dark clouds which constittues the object of the present study.

Recent calculations for the photodetachment rates in C$_{2}$H$^{-}$,
C$_{4}$H$^{-}$, and C$_{6}$H$^{-}$ \cite{Douguet2014} have further shown
that, for the cases of C$_{2}$H$^{-}$, C$_{4}$H$^{-}$, and C$_{6}$H$^{-}$ the
agreement with existing experiments \cite{Kumar2013} varies from system to system 
and the cross section behaviour at threshold strongly depends on dipole strength effects.
They have later analysed explicitly the REA process in the same series of  linear anions by 
further including a model treatment of the indirect processes \cite{Douguet2015}.They 
found that the REA rates for  C$_{2}$H$^{-}$ species were practically negligible and that
those for the C$_{4}$H$^{-}$ member were also rather small. As done before, by using the 
initials of all the authors in that paper, here we shall call this model the Douguet-Fonseca-Raoult-Dulieu-Orel-Kokoouline (DFRDOK)  modelling of the REA processes of the present study.

 In conclusion, the existing models for generating REA cross sections
and rates indicate the processes to come from combined contributions of direct
and indirect electron attachment paths that  can generate  effective
competition with the autodetachment and photodetachment channels, 
thereby yielding significant rates for the anion-forming process.
As briefly outlined above,  the HO model includes  no scattering calculations, 
while the CSGG approach carries out realistic scattering 
calculations but includes only indirectly non-adiabatic couplings with the internal
 vibrational modes. Such effects are modelled explicitly in the DFRDOK 
treatment for small molecules. They however found  these effects to be inefficient 
for the two smallest members of the polyyne chains,
thus making that channel essentially non-existent for the C$_{2}$H$^{-}$ and rather
marginal for the C$_{4}$H$^{-}$, the only two systems analysed by that model \cite{Douguet2015} .

 In order to gain further insight on the global effects that
the REA rates, as given by the existing calculations, might have on producing 
agreement with the  A/N ratios experimentally observed, we shall
carry out in the following section a more comprehensive computational analysis of the
anions' evolutions within an extended kinetic treatment of the network of chemical
reactions generally considered to be relevant for their formation in the core of molecular
clouds.

\section{Role of REA rates in chemical evolution calculations}

 The recent observations of \cite{Cordiner2013} have given us
specific indications for the C$_{4}$H/C$_{4}$H$^{-}$ and
C$_{6}$H/C$_{6}$H$^{-}$ ratios over seven nearby galactic star-forming cores
and molecular clouds.
The new measurements include four sources in which no molecular anions had been
previously detected and indicate that for the C$_{6}$H the observed A/N ratio the 
value is $\ge$1\% in every source, providing a mean value of 3\%.
On the other hand, for C$_{4}$H the detected A/N ratio value turned out to
be much smaller, i.e. $1.2\pm0.4\times10^{-5}$.
Furthermore, the earlier measurements for the C$_{8}$H$^{-}$/C$_{8}$H ratio in
the galactic molecular source TMC-1 \cite{Brunken2007} and in the
circumstellar envelope IRC+10216 \cite{Remijan2007} indicated the
ratio of the total column densities to be between $\sim$5\% and $\sim$3.8\%.

 It therefore becomes relevant to revisit the current chemical 
models for these quantities and to see how important the REA path could  be
in providing realistic A/N ratios in molecular clouds from 
evolutionary  calculations.
As a preliminary test, we have employed a one-zone time-dependent scenario
representing a dark cloud core (described e.g. in \cite{Wakelam2008}).
The selected model of the cloud evolves from a chosen set of initial
abundances at constant temperature, number density, cosmic-ray flux, and gas
visual extinction.
We employ the rates for the chemical database KIDA\footnote{http://kida.obs.u-bordeaux1.fr/}, using their most updated available set of reactions, i.e. \verb+kida.uva.2014+ \cite{Wakelam2015}.

 This set includes anions for C$_{n}$H$^{-}$ ($n=4,6,8$) which take
part in initial mutual neutralization reactions, photodetachment, associative
detachment, anion-neutral reactions, and REA processes, as described in KIDA. 
When we include all the reactions within a temperature range that is realistic for our model ($T_{gas}=10$~K), 
the total number of species selected was 486, with a total number of reactions up to 6962.
The only difference we adopthere  is to take REA reactions rate coefficient for ($n=4,6,8$) anions formation
 from the recent scattering model termed CSGG in our present study \cite{Carelli2013}, rather than their simpler Langevin-type estimates usually included in the recent databases.
The present calculations employ the open-source code \textsc{KROME}\footnote{\url{http://kromepackage.org/}} \cite{Grassi2014} to solve the time-dependent system of differential equations. In order to ensure numerical accuracy we set relative and absolute tolerances for the numerical solver to $10^{-12}$ and $10^{-40}$ respectively.

 We also selected the initial abundances for the different species, as
reported in earlier work following Tab.~1 of \cite{Cordiner2013}, but we will also show the results by employing model EA2 from \cite{Wakelam2008}, that takes into account more metals, i.e. He, F, Na, Mg, Si, P, Cl, and Fe.
Furthermore, to guarantee global cloud neutrality in the initial conditions, we choose the number density
of the free electrons to be $n_{\rm e}=\sum_i n_i$ by running the index $i$ over
all the positive ions present in the network, while during the system evolution the electron abundance is solved by using the same set of differential equations employed for the other chemical species, that guarantees global charge neutrality by construction.
With these initial conditions the reactions including C$_n$H$^-$ are 236, and a sample of them is listed by Tab.~1. The mutual neutralization reactions of line 5 in the Table also include the C$^+$ partner.

 The other parameters of our calculations are the gas temperature
$T_{gas}=10$~K, the visual extinction \mbox{$A_{V}=10$~mag}, the cosmic-ray ionization
rate $\zeta=1.3\times10^{-17}$~s$^{-1}$ and the dust/gas mass
ratio $D=0$, i.e. a dust-free model. We are aware that the latter assumption could affect the chemical evolution, since dust (and PAHs) may have a large impact on the gas charge \cite{Tielens2005} and enable surface chemistry \cite{Hocuk2015}; However, since we are interested in the role of the new REA calculations, we consider this approximation satisfactory. To reduce the uncertainties of our analysis this model employs a chemical network taken from KIDA, which includes for the above reactions  the most recently available reaction rate coefficients. On the other hand, the calculations reported in  \cite{Cordiner2013} are based on the earlier data from UMIST06\footnote{\url{http://udfa.ajmarkwick.net/}}. It should be also mentioned here
that to vary the gas temperature in each run of our model is likely to affect the final abundances, although the chosen value is considered the most emblematic temperature value when modeling a cold core in a Dark Molecular Cloud. Hence, we shall restrict our analysis to the chosen value of T.

\begin{table}
\begin{tabular}{llll}
\hline
1. & C$_n$H + e$^-$		& $\rightarrow$	& C$_n$H$^-$ + $h\nu$ \\
2. & O + C$_n$H$^-$		& $\rightarrow$	& CO + C$_{(n-1)}$H + e$^-$ \\
3. & O + C$_n$H$^-$		& $\rightarrow$	& CO + C$_{(n-1)}$H$^-$ \\
4. & O + C$_{(n+1)}$H$^-$	& $\rightarrow$	& CO + C$_{n}$H$^-$ \\
5. & X$^+$ + C$_{n}$H$^-$	& $\rightarrow$	& X + C$_{n}$H \\
6. & H + C$_{(n+2)}$H$^-$	& $\rightarrow$	&   C$_2$H + C$_{n}$H$^-$\\
\hline
7. & HCO$^+$ + C$_{n}$H$^-$	& $\rightarrow$	& HCO + C$_{n}$H\\
8. & HCO$^+$ + C$_{n}$H$^-$	& $\rightarrow$	& H + CO + C$_{n}$H \\
9. & C + C$_{n}$H$^-$	& $\rightarrow$	& C$_{(n+1)}$H + e$^-$ \\
10. & H$_3$O$^+$ + C$_{n}$H$^-$	& $\rightarrow$	& H + H$_2$O + C$_{n}$H \\
11. & N + C$_{n}$H$^-$	& $\rightarrow$	&   C$_{(n-3)}$H + C$_3$N$^-$\\
12. & N + C$_{n}$H$^-$	& $\rightarrow$	&   C$_{(n-1)}$H + CN$^-$\\
\hline
\end{tabular}
\caption{A sample of the key reactions for the formation/destruction of C$_n$H$^-$ from KIDA database included in the present study, except reaction~1 that is also from \cite{Carelli2013}. The species X represents a generic metal (X = Mg, Na, Fe, C). Note that the reactions are not necessary equally effective for $n=4,6,8$, e.g reaction~6 is only important for $n=8$. Also, the reactions are not sorted by their relative effectiveness, however 7-12 are less important than 1-6 for our discussion. See the main text for more details.}
\end{table}

 We have run our calculations using the REA rates we had reported in
our earlier CSGG modelling \cite{Carelli2013} and which turned out to be rather close
to those produced by the earlier estimates in the HO modelling  \cite{Herbst2008}, albeit
  uniformly smaller than the latter.
A detailed, albeit simpler comparison of the  REA rates produced by the existing calculations has already
been done by us \cite{Carelli2013} and will not be repeated here. What we shall analyse instead are the effects from scaling the computed REA rates on the A/N ratios of the polyyne anions available from experiments, as well as the effects on  their absolute abundances.

\section{Results and discussions}

\begin{figure}
\centering
\includegraphics[width=0.5\textwidth]{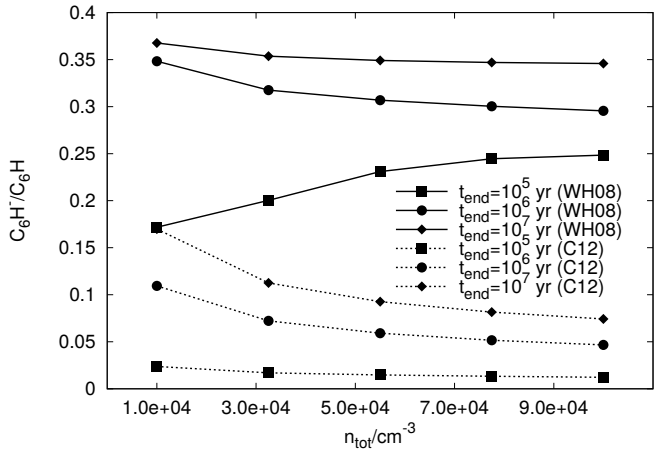}
\caption{Computed A/N ratios for the C$_{6}$H$^{-}$ anion following the
evolutionary model described in the main text.
The curves are referring to different times in the evolutionary modelling, i.e. $t_{end}=10^5$~s (squares), $t_{end}=10^6$~s (circles), and $t_{end}=10^7$~s (diamonds),   and initial conditions, i.e. from \cite{Cordiner2012} (C12, dashed) and \cite{Wakelam2008} (WH08, solid). The figure reports the changes in the A/N ratio as a function of the total density over the same range discussed by \cite{Cordiner2013}.}
\label{fig01}
\end{figure}

In the survey on A/N ratios reported by \cite{Cordiner2013} it was also found by their simulations that,
in the case of the C$_{6}$H$^{-}$ anion, there was a positive correlation between the value
of that ratio and the total number density. Their Fig.~5 indicated, in fact, that the ratio
varied between $\sim$2.5\%  and $\sim$4.4\% when the hydrogen molecule's density was varied by about
 one order of magnitude.
Similar calculations were carried out by us using the present model with the chemical set-up previously described (see Fig.~\ref{fig01}): the general behaviour is seen to  depend on the time evolution ($t_{end}$) and on the initial conditions employed. In the case of \cite{Cordiner2012}, given by the initial conditions (C12), we find a negative dependence of the A/N ratio with the total density that, as expected, converges to a constant value for long evolutions ($t_{end}=10^8$~yr). In the case of \cite{Wakelam2008}, given by the  initial conditions (WH08), we note that the positive trend is found for $t_{end}=10^5$~yr, but this dependence is cancelled at later evolutionary stages ($t_{end}\geq10^6$~yr). 

The difference with the results found by \cite{Cordiner2012} is due to the different set-up employed by us and the more recent rates and chemical network values used within our modelling. We note here, in fact, that the initial conditions play a key role in determining the comparison with the observed quantities, especially when more metals are present (WH08 case). The effects due to the presence of the dust are not taken into account but, as mentioned in the previous section, they could modify the behaviour found in all models.

\begin{figure*}
\centering
\includegraphics[width=1.1\textwidth]{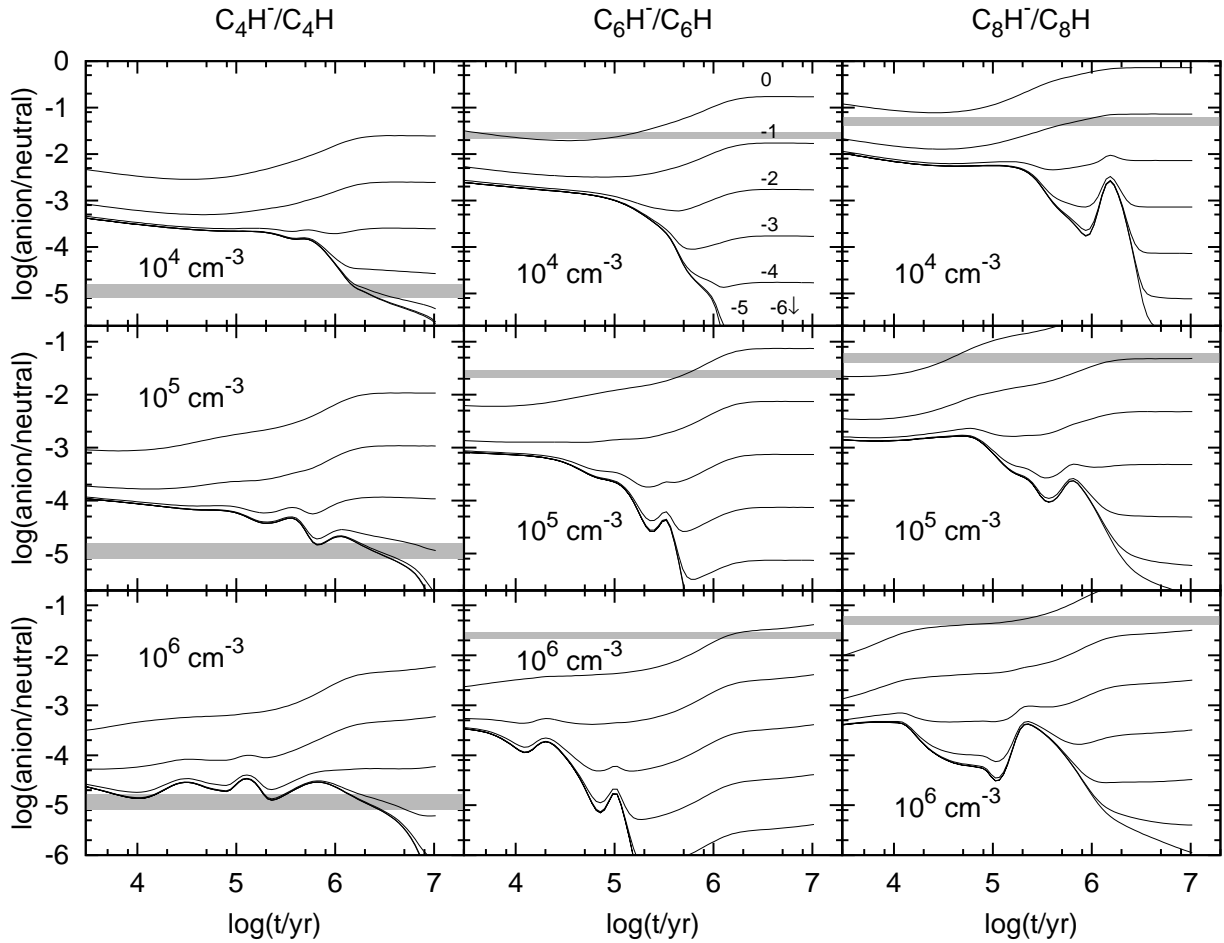}
\caption{Computed A/N ratios for C$_{n}$H$^{-}$ ($n=4,6,8$), obtained by following the
evolutionary model described in the main text. In each of the panels the ratios have been artificially scaled
over six orders of magnitude from top to bottom, see as a reference the labels with $\log(n_{tot}/{\rm cm}^{-3})$ in the second panel, top row, where the powers of ten used in the scaling for all panels are explicitly reported.The scalings have been obtained by varying of the same factors the REA rates employed in the calculations. The A/N experimental values (with errors) are given by the grey-shadowed areas in each panel. Note that observed values are valid for TMC-1~CP, that has density of $10^{4}$~cm$^{-3}$ and an age of $10^5$~yr. The choice of the varying total densities indicated in each panel are explained in the main text when discussing this figure. The experimental values with their errors are from \cite{Cordiner2013} and \cite{Brunken2007}. Initial conditions are from \cite{Cordiner2012}. The second panel, top row, is also discussed in Fig.~\ref{fig:cfrC6H}.}
\label{fig02}
\end{figure*}

\begin{figure*}
\centering
\includegraphics[width=.9\textwidth]{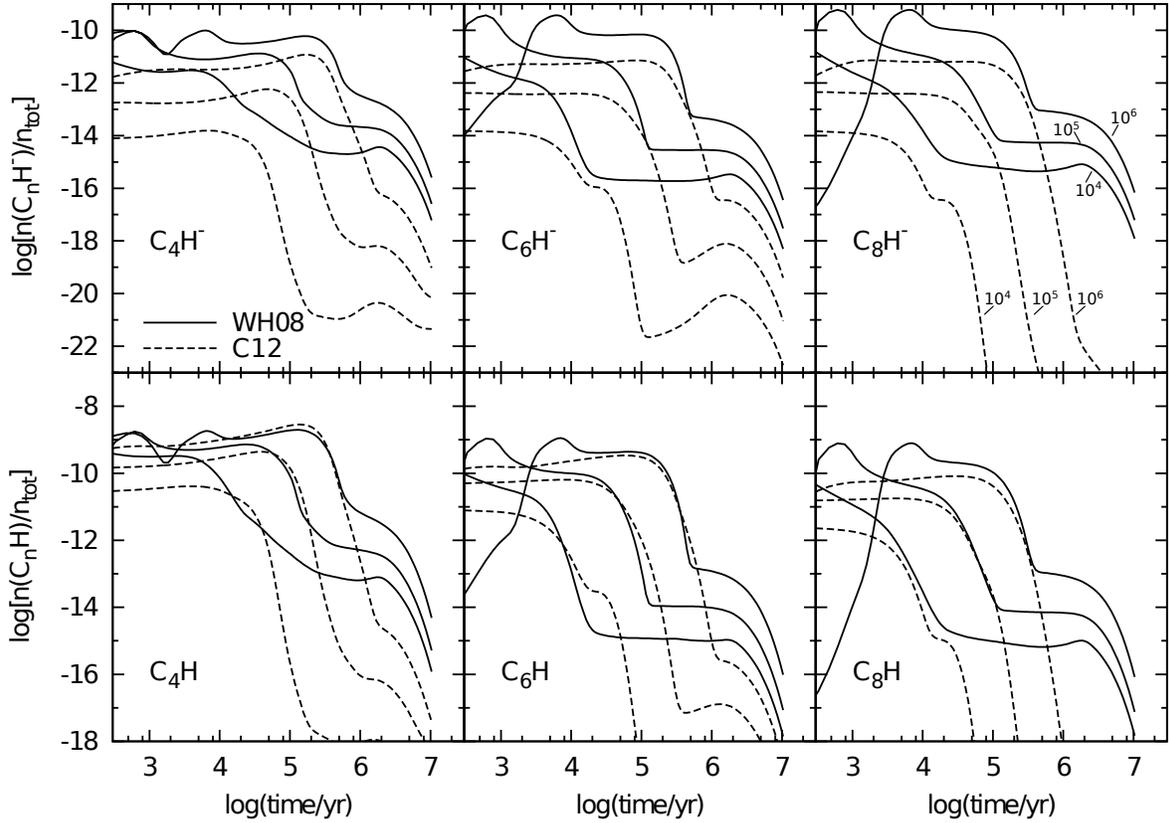}
\caption{Computed time evolutions for C$_{n}$H$^{-}$ ($n=4,6,8$) and their anions, following the
evolutionary modeling described in the main text. In each panel we report the computed evolutions using different choices for the initial conditions:  from \cite{Wakelam2008}, labelled  as WH08, and from \cite{Cordiner2012}, labelled  as C12. The labels in the third panel, top row, indicate the various values for the total cloud density and are omitted in the other panels because they are the same .}
\label{fig03}
\end{figure*}

We have carried out a further, and more extensive, test on the dependence of the A/N ratios on the total density (i.e. H$_2$ density) and evolution time for C$_{n}$H$^{-}$ ($n=4,6,8$). In particular, we have run simulations based on three different choices of such densities (see Fig.~\ref{fig02}): the low-density values typical of a quiescent cloud ($10^4$~cm$^{-3}$, upper panels), the middle-range densities relevant for the star-forming regions ($10^5$~cm$^{-3}$, middle panels), and the high-density values associated with protostars (lower panels, $10^6$~cm$^{-3}$).
 Analogously, in Fig.~\ref{fig03} we report the absolute evolutions of neutral and corresponding anion species for different densities and initial conditions that show a behaviour similar to the one found by \cite{Cordiner2012}.

 When  we examine Fig.~\ref{fig02}, we see there that the ratio for the C$_{4}$H$^{-}$  matches the experimental ratios only when the REA value, initially taken from the  calculations of the CSGG or OH models  and included in all the simulations  of the cloud shown  in the three panels of that figure,  is reduced by nearly six orders of magnitude. This indicates that, within the quality of our chemical network, the REA path to the formation of that anion  must make only  negligible contributions to the overall  rates  of its formation. 
We further see that  when the macroscopic system has reached later stages in the time of its evolution, sometime even later than the estimated cloud age, the scaling can be reduced to only about three orders of magnitude to match experimental values.
Furthermore, the analysis of the evolutionary models indicates that at low-densities it is chiefly the reaction \mbox{O + C$_5$H$^-\rightarrow$ CO + C$_4$H$^-$} that supports the anions formation while there are  no significant contributions to its formation coming from the REA process.

It is important to note , especially at later times in the cloud's evolution, that the results of our simulations reported in the center- and right-columns of panels of Fig.~\ref{fig02} for the behaviour for the other two  members of the present series of anions indicate a significant role of the REA mechanism ( from the CSGG model  which we have included  in our study) to match the experimental values given by the shadowed areas in all the panels.
In fact, the larger anions  show that the efficiency of the REA channel to their formations produced by the present CSGG theory is matching the experimental observations: they are in fact reproduced  by using less than an order of magnitude reduction of the theoretical values.
\noindent It is also interesting to note that the theoretical values we have employed are orders of magnitude smaller than the Langevin-type estimates usually included in the databases: for the C$_{4}$H$^{-}$ case the corresponding value reported is $1.10\times10^{-8}\left(T_{gas}/300~{\rm K}\right)^{-0.5}$. Likewise, for the cases of C$_{6}$H$^{-}$ and of C$_{8}$H$^{-}$ the Langevin estimates are also much larger than the ones provided by the CSGG theoretical modelling: $6\times10^{-8}\left(T_{gas}/300~{\rm K}\right)^{-0.5}$. It therefore follows that the values produced by the CSGG model  for the REA rates for  the longer chain  anions are much more realistic than the Langevin rates.They further show a temperature dependence, feature which is absent in  the latter approximation  \cite{Carelli2013}.

If we now compare the  evolutions of the absolute abundances, in Fig.~\ref{fig03}, for both the neutral compounds and their anions, we note that for ages of the cloud larger than $10^6$~yr the amount of both the neutral C$_{n}$H and of their anions reaches very low concentrations with respect to the total gas density, especially for the initial conditions  labelled~C12. This suggests that in the employed models the error carried by the A/N ratios after long evolutionary times could become larger than expected, since the uniformly small values of  their  absolute quantities naturally have a substantial effects on the errors of their ratios. It then follows that  the role played  by obtaining increasingly more accurate REA rates  becomes also  crucial in controlling the quality of their A/N ratios.

\begin{figure}
\centering
\includegraphics[width=.47\textwidth]{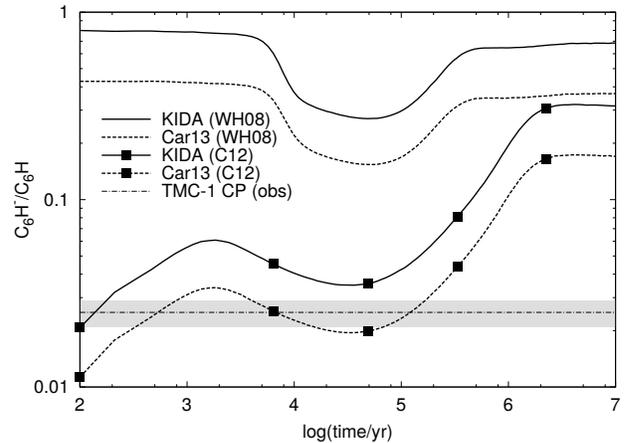}
\caption{Computed time evolutions of the A/N ratio  for C$_6$H, using different initial conditions, i.e. \cite{Cordiner2012} (C12, lines with squares) and \cite{Wakelam2008} (WH08, lines without squares), and also employing different electron attachment (REA)  rate coefficients. The latter are from \cite{Carelli2013} (Car13, dashed lines) and from the KIDA database (KIDA, solid lines). The observational  values from \cite{Cordiner2013} for TMC-1~CP  are also reported (dash-dotted line) in the figure,  together with their errors (grey-shadowed area).}
\label{fig:cfrC6H}
\end{figure}

Another indication of the importance of including the most realistic REA rates is provided, for the case of the 
C$_{6}$H$^{-}$ anion's A/N ratio, by the evolutionary comparison reported in Fig.~\ref{fig:cfrC6H}, where the low density case associated with the TMC-1~CP conditions of a quiescent cloud \cite{Cordiner2013} are also employed in our calculations. We note here that, when we use the  initial conditions employed by \cite{Cordiner2012} (C12),while including  the REA rates from the CSGG model  \cite{Carelli2013}, the results labelled  (Car13) match the observational data for ages that are compatible with that of the TMC-1~CP cloud, i.e. $\sim10^5$~yr. On the other hand, the A/N ratio given by the present modelling when using the REA rate value given by  the KIDA database is larger  than the observed one  for nearly the whole evolution time. As mentioned in the previous cases, adding more metals to the initial conditions (i.e. using the WH08 prescription) affects the global behaviour of the A/N ratio, as clearly indicated by the upper curves of the present  Fig.~\ref{fig:cfrC6H}.

The following general considerations could be made by looking at all the figures'
data presented thus far,also keeping in mind the effects from the most efficient 
chemical reactions already mentioned earlier:

\begin{enumerate}[label={(\roman*)}]

 \item all the figures consistently show that a reduction in
size of the REA rates  within the modelling causes an almost linear reduction of
the A/N ratios to be compared with observations, in spite of the complexity of the
 chemical links which are active within each of the
networks that we have used in the evolutionary modeling.

 \item  When the REA rates are reduced by more than five-six orders of
magnitude, the ``humps'' shown by the time evolution curves, around time 
 evolution values between 10$^{5}$~yr and 10$^{6}$~yr,  were found to be linked
 to the following chemical formation channels, an effect occurring independently
 of the initial conditions:

\begin{equation}\label{eqn:reaOC}
 \mO + \mC_{n+1}\mH^{-} \rightarrow \mC\mO + \mC_{n}\mH^{-}\,,
\end{equation}
for $n=4,6$ at low densities, and $n=4,6,8$ at high densities, while
\begin{equation}
 \mH + \mC_{n+2}\mH^{-} \rightarrow \mC_2\mH + \mC_{n}\mH^{-}\,,
\end{equation}
for $n=8$ at low densities. These reactions turn out to be the main anion formation channels at the times when the REA processes become negligible. Thus, for this reason they represent the lower limits of the anion formation efficiency. Note that both at low and high densities, when REAs are still significant processes in our modelling, the attachment efficiency is always larger than reaction~(\ref{eqn:reaOC}), except for $n=4$, where reaction~(\ref{eqn:reaOC}) dominates at low densities.

\item  In the database employed here  the reactions: \mbox{O + C$_{n}$H$^{-}$} and \mbox{O + C$_{n}$H} have different efficiencies in yielding their products, so that when the  REA anion formation rates are very small the chemical kinetics of the anions is faster than the analogous chemistry for neutrals.
As expected, when the REA rates are fully operative, or reduced by less than five order of magnitude, the A/N ratio increases because the REAs reduce the efficiency of the two chemical processes mentioned above. 
It is important to note here that these reaction rates are still given by the databases as the simple  Langevin-type reactions  \cite{Harada2008}, and therefore  more accurate calculations, or more modern experiments, on such reactions would help to better clarify this issue.

\item  The uniform drops of the absolute densities shown in our model by the data in Fig.~\ref{fig:cfrC6H}  for
C$_{n}$H$^{-}$ after time values around 10$^{5}$~yr are due to the following contributing reactions:

\begin{equation}
 \mO + \mC_{n}\mH^{-} \rightarrow \mC\mO + \mC_{n-1}\mH^{-}\,,
\end{equation}
this being true when using the C12 initial conditions, while for the different choices of  WH08 we found that
\begin{equation}
 \mC_{n}\mH^{-} + {\rm Mg}^{+} \rightarrow \mC_{n}\mH + {\rm Mg}
\end{equation}
\begin{equation}
 \mC_{n}\mH^{-} + {\rm Na}^{+} \rightarrow \mC_{n}\mH + {\rm Na}
\end{equation}
\begin{equation}
 \mC_{n}\mH^{-} + {\rm Fe}^{+} \rightarrow \mC_{n}\mH + {\rm Fe}
\end{equation}
 become an important chemical paths to anion
destruction, since more metals are present in the selected cloud's conditions.
This result is similar to what we have recently found on CN$^{-}$ formation \cite{Satta2015}, where we have shown  that when the electron-driven processes become slow and ineffective in forming anions,
 then other chemical reactions take over and  become dominant with respect to the REA mechanism. 
 Furthermore, this  inefficiency in the case of the CN$^{-}$ formation was also shown by  the accurate calculations on the REA path for the same molecule discussed in \cite{Douguet2015}.

\item  When looking at the C$_{4}$H$^{-}$/C$_{4}$H evolution over
time, we see that to match the A/N experimental value  \cite{Cordiner2013} requires 
 an electron attachment REA rate which should be  $\leq10^{-12}$~cm$^{3}$~s$^{-1}$.
This value, at 10~K, is at least four orders of magnitude smaller than that
estimated by the existing calculations with the CSGG modelling \cite{Carelli2013},
 which we have employed here, and  even smaller than the larger value from the HO modelling  \cite{Herbst2008}, which was  also around 10$^{-8}$~cm$^{3}$s$^{-1}$. 
On the other hand, the more recent results from the DFRDOK model \cite{Douguet2015} include an approximate form  of non-adiabatic coupling between the impinging electrons and the molecular 
vibrations  \cite{Douguet2015}.They  found for this system an REA rate of  formation
 around $10^{-16}$~cm$^{3}$~s$^{-1}$  at 30~K. In other words, both the  latest modelling  for the REA process, and the present modelling that examines several scalings of that rate, suggest that the REA path to C$_{4}$H$^{-}$ formation in the cores of cold molecular clouds  should be a more  inefficient path  than previously indicated. 

\item  In the case of the C$_{6}$H$^{-}$/C$_{6}$H ratio, the recent experiments by \cite{Cordiner2013} indicate an A/N mean value over all their sources of 3.10\%.
From the present numerical experiments of Fig.~\ref{fig02}, we see that to match their estimates  corresponds to employing in our modelling an REA rate of $5\times10^{-9}$~cm$^{3}$~s$^{-1}$, which is  now only about one order of magnitude lower than the previously calculated values obtained by the CSGG dynamical modelling \cite{Carelli2013,Herbst2008}. Given the uncertainties that we know to exist among several of the chemical rates reported in our employed database (KIDA), one can consider the discrepancy between estimated and computed REA formation rates for this member of the series to be within the
expected error bars of the present runs, where the agreement with observations is bested around $10^{5}$~yr.

\item  The results of Fig.~\ref{fig02} for the A/N ratio involving the longest member of the present series, 
the C$_{8}$H$^{-}$/C$_{8}$H ratio, further indicate that the experimental values between 3.8\% and
5\% is accurately given when using an REA attachment rate, at 10~K, of about $10^{-8}$~cm$^{3}$~s$^{-1}$. The CSGG computational  estimate used in our present study , at 10~K, is \mbox{$3\times10^{-7}$~cm$^{3}$~s$^{-1}$}, i.e. only  about one order of magnitude larger. This indicates once more that the quantum dynamical models of reference \cite{Carelli2013} become increasingly more reliable for the larger members of the series, while overestimating REA rates for the smallest members.
\end{enumerate}

 In all cases, the large A/N ratio's drops in value as time increases aredue to the different chemical efficacy  of the reactions  \mbox{O + C$_{n}$H$^{-}$} and \mbox{O + C$_{n}$H}, where the latter  generally exhibits smaller  rate coefficients. The slower chemical processes reduce the abundance of the neutrals less efficiently than that of  the anions, hence causing the latter to disappear in the dark clouds more rapidly than the neutrals. This behaviour induces a  reduction of  their A/N ratios as the evolutionary times of the clouds increase.
 Analogously, when the A/N ratios increase, the presence of the REA processes offsets  the effects of the two reactions with atomic oxygen. They become more significant in controlling anions' formation paths.
 Our present evolutionary study indicates that the attachment
rates existing in the current literature are uniformly larger than those which 
can match the experimental A/N ratios at 10~K within rather sophisticated
 chemical networks, as those we have  used to describe the molecular cloud's evolutions in time.
Finally, we should also note here that we have  tested the robustness of our evolutionary 
model by running it at at least one higher temperature, i.e. at 30~K, and found essentially no changes in all the parameters produced by the model.

Our earlier scattering calculations using the CSGG model \cite{Carelli2013}, which we have
employed here within a fairly large chemical network of reactions, had
already suggested that the computed REA values using that model were to be considered
as upper bounds to the true rates,  since no estimates for the 
autodetachment channels were included explicitely.
The present scaling experiments on the A/N evolution in dark molecular clouds
confirm that, for the case of the smallest member of the series examined, the actual REA rates 
have  to be smaller than those suggested by calculations: for C$_{4}$H$^{-}$) the rates required to match experimental A/N ratios should be  about four orders of magnitude smaller.

 On the other hand, as the length of the carbon chain increases, e.g. for
C$_{6}$H$^{-}$ and C$_{8}$H$^{-}$, the CSGG model calculations indicate that, to obtain
realistic A/N ratios within the evolutionary model of a dark molecular clouds,
our calculated REA rates  are only slightly larger: given the general uncertainties 
for many of the chemical rates contained in the chosen database, the calculated electron
attachment rates are closer to reality for the latter molecules than for the
case of  the C$_{4}$H$^{-}$ anion. As a possible explanation for these differences, 
it is worth mentioning at this point that C$_{4}$H is the only partner molecule in the series
 which does not possess a critical value for its dipole moment  \cite{Carelli2013} and therefore would
 not have the dipole-driven metastable states which can contribute to a more efficient REA
mechanism, as we have explained earlier in the Introduction.
Since the REA rates are expected to increase with molecular complexity (e.g. for
larger polyynes), it seems reasonable to surmise  that rates obtained using the CSGG model 
for the longer C-bearing chains of these molecular systems
\cite{Carelli2013} would be likely to  match  even better future possible
observations of A/N ratios in  such longer chains.  

\section {Present Conclusions}\label{sect:conclusions}

 The present numerical experiments indicate that the
efficiency of the electron attachment processes, and therefore their role in
forming anions of the linear polyynes discussed in the present work,
varies a great deal within the members of the  series and is strongly linked to 
the chemical features of each specific molecular structure.

As a specific example, we have shown that  the subcritical dipole moment in C$_{4}$H$^{-}$, 
and its fairly small number of vibrational degrees of freedom, can be  taken as  chemical indicators and physical reasons for its markedly smaller REA rates at low temperatures 
in comparison with those for C$_{6}$H$^{-}$ and C$_{8}$H$^{-}$, both longer  molecules 
with larger permanent dipoles.

 Our computational modelling of REA processes, the CSGG model  based on scattering calculations
 \cite{Carelli2013}, has been proposed a few years ago and  
 should  provide upper bounds to the  actual rates of attachment since no direct 
treatment of autodetachment channels  was included in it. The inclusion of such effects 
for the two smallest members of the  polyyne series in the later work of \cite{Douguet2015} has  shown that the corresponding REA rates swould be much smaller than expected because of the efficient 
competition provided by autodetachment channels.

The  direct testing of our CSGG scattering modelling, once we employ it within the evolutionary chemical model discussed in the present  study, is therefore showing that REA channels to anion formation are of 
negligible importance for producing C$_{4}$H$^{-}$ molecules. One should note here that the above 
is valid within the general quality of the chemical network which we have employed, and which is the most up-to-date selection of chemical rates currently available for such evolutionary models. 

The same REA mechanism however contributes to the final densities of C$_{6}$H$^{-}$ and C$_{8}$H$^{-}$ 
anions with rates that are two-to-three orders of magnitude larger than that of the shorter chain of polyynes. Such rates are now very close to the values proposed by the CSGG and the HO 
 models \cite{Carelli2013,Herbst2008}, thereby validating their estimates for these specific  rates. 

The various features of the chemical network employed in the present modelling also suggest that the evolutionary age (and then the initial conditions) of the object under consideration can play a significant 
role since the chemical reactions that produce both the neutral members and their anionic counterparts
become less efficient over longer time spans and consistently reduce the absolute densities of both members in the A/N ratios we are discussing. As a consequence of it, the expected error bars for such ratios would increase and thus make them  likely to be affected by much larger errors.

In conclusion, the present detailed study on the effects of REA  rate sizes within a fairly large set of chemical reactions,specific for the members of the C-bearing linear polyynes, indicates that 
the values of their A/N ratios can vary a great deal within the members of that series and that their specific chemical features directly influence the relative importance of the REA formation of their stable anions. The desire is that such quantitative analysis of the comparison between observational data and computed data from evolutionary models can help our community to better understand the role of the REA mechanism within the larger scene of anion formation in  Molecular Clouds.


\section*{Acknowledgements}

 The financial support from the Austrian Research Agency FWF through Project no
Project P27047-N20 and from the COMIQ Research Training Network funded by the E.U.
 is gratefully acknowledged. TG acknowledges the Centre for Star and Planet Formation funded
by the Danish National Research Foundation. 


\section*{References}

\bibliography{jchem}

\end{document}